\documentclass[aps,prl,showpacs,twocolumn,groupedaddress]{revtex4}

\usepackage{graphicx,color}
\usepackage{bm, amsmath, amssymb}

\begin{document}

\title{ Vortex Dynamics: Quantum versus Classical Regimes }

\author{L. Thompson$^{1,2}$ and P.C.E. Stamp$^{1,3}$}

\affiliation{$^{1}$Department of Physics \& Astronomy, University
of British Columbia, Vancouver, BC, Canada V6T 1Z1 \\
$^2$Department of Physics, Massachusetts Institute of Technology,
Cambridge, MA 02139\\
$^{3}$Pacific Institute of Theoretical Physics, Vancouver, BC V6T
1Z1, Canada}

\begin{abstract}

For many years the classical Hall-Vinen-Iordanski (HVI) equation has
been used to analyse vortex dynamics in superfluids. Here we discuss
the extension of the theory of vortex dynamics to the quantum
regime, in which the characteristic vortex frequency is higher than
the temperature. At the same time we justify, in the low-frequency
classical regime, the use of the HVI equation, provided an inertial
mass term and a noise fluctuation term are added to it. The
crossover to the quantum regime is discussed, and an intuitive
picture is given of the vortex dynamics, which in general is
described by 2 equations (one for the vortex coordinate, and one for
its quantum fluctuations); we also discuss the simple equation of
motion found in the extreme quantum regime.

\end{abstract}

\maketitle

\section{Introduction}

Very soon after the discovery of quantum vortices in superfluid
$^4$He by Vinen \cite{vinen56}, an equation of motion for a vortex
was proposed by Hall and Vinen\cite{HallV}; a few years later
Iordanskii\cite{iord} added an extra term, to produce what is
commonly known as the Hall-Vinen-Iordanskii (HVI) equation of
motion. The HVI equation, occasionally supplemented by an inertial
mass term and by a noise fluctuation term, has been used in the last
60 years to analyze thousands of experiments in superfluids and
superconductors. However it has been controversial, and for the last
20 years a strenuous debate has been going on over its validity. Key
questions concern the value of the vortex effective mass $M_v$
(estimates range from zero to infinity \cite{anglin07}) and the
vortex-quasiparticle coupling coefficients $D_o(T)$,
$D_o^{\prime}(T)$. Indeed, Thouless et. al. \cite{TAN96} find
$D_o^{\prime}(T) = 0$ for all $T$; and scattering
analyses\cite{sonin97,iord, fetter64,wexthou98,stone00} give various
different results for $D_o^{\prime}(T)$.

In what follows we wish to shift the focus of this discussion. In
our view, the key question is how to give a proper
quantum-mechanical description of the vortex dynamics. This requires
two equations of motion, one for a vortex 'centre of mass'
coordinate, and the other for the quantum fluctuations around this
coordinate\cite{TS12}. A key parameter in the theory is the
dimensionless ratio $\tilde{\Omega} = \hbar \Omega/kT$, where $T$ is
the temperature, and $\Omega$ the characteristic frequency of the
vortex dynamics. In the classical limit, where $\tilde{\Omega}
\rightarrow 0$, one actually recovers the HVI equation with added
inertial and noise terms\cite{TS12}. While key questions remain
(notably the role of boundaries and normal fluid velocity
\cite{ThV01}, and the calculation of the effective
mass\cite{anglin07}), we believe that the main points at issue have
now been settled in the classical regime.

However the quantum regime remains relatively unexplored. It
possesses a number of fascinating features, both theoretical and
experimental. The theory gives an illuminating picture of
vortex-quasiparticle interactions, classical analyses of which
interactions have been subtle and controversial, because of the
long-range nature of the interaction, and the difficulty of
accounting for vortex 'recoil' in the scattering. A quantum analysis
immediately makes clear that a key feature of the interaction (which
is properly described as an interaction between quantum soliton and
quasiparticle field excitations in $2+1$ dimensions), is the
distortion of the quasiparticle part of the superfluid wave-function
by the vortex 'zero mode' part of this wave-function. On the
experimental side the results suggest new kinds of experiment, and
show that most previous experiments have been in the classical
regime.

In this paper we present a more intuitive picture of the quantum
regime for vortices in a Bose superfluid, highlighting the new
features. On the theoretical side we first discuss the main features
of a quantum description of the vortex, and of vortex-quasiparticle
interactions. We then briefly summarize the resulting equations of
motion for the vortex, and give an intuitive picture of them in real
time (as opposed to frequency space). Almost all details of the
calculations are eschewed - they are quite lengthy and will appear
elsewhere\cite{TSlong}. We then very briefly discuss how one might
experimentally probe the quantum regime in a Bose superfluid. We
emphasize that this paper is about vortices in neutral Bose
superfluids - the problem of vortex dynamics in Fermi superfluids
(both charged and neutral) is rather different (and is, we believe,
still open).


\section{Quantum-Mechanical description of a superfluid vortex}

In a classical description, a superfluid vortex is described by its
position ${\bf R}_v(t)$ (and time derivatives of this position), and
its dynamics are specified by an equation of motion for ${\bf
R}_v(t)$.

Consider now a quantum description of a superfluid with a single
vortex in it. One begins with the $N$-particle wave-function
$\Psi(\{ {\bf r}_j \}) = \langle \{ {\bf r}_j \} | \Psi \rangle$,
where $j = 1,2,....N$ and $|{\bf r}_j \rangle$ is a position state
for the $j$-th particle; or, equivalently, from the $N$-particle
density matrix ${\bm \rho}_N(\{ {\bf r}_j \}, \{ {\bf r'}_j \}) =
\langle \{ {\bf r}_j \} |\hat{\bm \rho}_N | \{ {\bf r'}_j \}
\rangle$, where $\hat{\bm \rho}_N =  |\Psi \rangle \langle \Psi |$.
The state-vector $|\Psi \rangle$ and the wave-function $\Psi(\{ {\bf
r}_j \})$ are assumed to satisfy Bose symmetrization over
permutations of the particles. We now add a vortex soliton to the
system, with circulation $q_v {\bm \kappa}$, where $q_v = \pm 1, \pm
2, ...$, and assume that in the  $N$-particle wave-function the
vortex node is  at a point ${\bf R}(t)$ in the plane. Both the
$N$-particle wave-function and the total density matrix $\hat {\bm
\rho}_N$ must then depend explicitly on the parameter ${\bf R}(t)$.
One is now free to make a change of variables to a set $\{ {\bf R};
{\bf q}_k \}$ of `collective coordinates' \cite{Raja}, with $k =
1,2,....N-1$, wherein the vortex coordinate ${\bf R}(t)$ is isolated
from the remaining $N-1$ coordinates $\{ {\bf q}_k \}$. This vortex
coordinate, formerly just a parameter in the wave-function, is now
elevated to the status of a quantum variable associated with a state
vector $| {\bf R} \rangle$ and with a `zero mode', which we discuss
below. All other degrees of freedom must now be properly
orthogonalized, both to the zero mode and to each other \cite{Raja}.
The corresponding $N$-particle density matrix is written ${\bm
\rho}_N(\{ {\bf R}, {\bf q}_k \}; \{ {\bf R'}, {\bf q'}_k \}) =
\langle \{ {\bf R}; {\bf q}_j \} |\hat {\bm \rho}_N | \{ {\bf R'},
{\bf q'}_j \} \rangle$. We also define a vortex reduced density
matrix by integrating out the $\{ {\bf q}_k \}$:
\begin{eqnarray}
\bar{\bm \rho}_v({\bf R}, {\bf R'}, t) &=& \mathrm{Tr}_{{\bf q}_k}
\;
{\bm \rho}_N(\{ {\bf R}, {\bf q}_k \}; \{ {\bf R'}, {\bf q}_k \}) \nonumber \\
& = & \prod_k \int d{\bf q}_k \;{\bm \rho}_N(\{ {\bf R}, {\bf q}_k
\}; \{ {\bf R'}, {\bf q}_k \})
 \label{rhoNred}
\end{eqnarray}
in terms of which all physical quantities associated with the vortex
may be defined and evaluated, provided we have sufficient
information about $\hat {\bm \rho}_N$.

One may now, by fairly well-established manoeuvres, derive a field
theory for the Bose-condensed system, starting from the 1-particle
reduced density matrix\cite{ons56}, denoted by ${\bm \rho}_1({\bf
r}, {\bf r'}, t)$. Note that this is a different object from the
vortex reduced density matrix we have defined above, and is obtained
in the usual way by integrating over the coordinates of all the
particles in the full density matrix except for one of them (in a
fully symmetrized way that takes account of the Bose statistics). As
is well known, one can characterize the result in terms of a quantum
phase field $\Phi({\bf r},t)$ and a density field $\rho({\bf r},t)$;
they are assumed to have Bose commutation relations (and in a
classical approximation, these fields are amalgamated into a
macroscopic wave-function $\psi({\bf r},t) \sim \rho({\bf r},t)e^{i
\Phi({\bf r},t)}$, and the commutation relations are dropped). It is
common in the literature to separate out a slowly-varying 'texture'
in the two fields, and write
\begin{align}
\Phi({\bf r},t) &= \Phi_s + \phi({\bf r},t)\\
\rho({\bf r},t) &= \rho + \eta({\bf r},t)
\end{align}
where the 'quasiparticle' variables $\phi({\bf r},t)$ and $\eta({\bf
r},t)$ describe fluctuations about the texture. However once we
introduce a vortex into the superfluid, we have to be a little more
careful. The vortex solution breaks the global translational
symmetry of the superfluid, leading to a new quantum mode associated
with this broken symmetry, the so-called vortex zero mode. Without
specifying a particular Hamiltonian $\mathcal H$ for the superfluid,
we can nevertheless say that the general equations that admit the
vortex are:
\begin{align}\label{vortex}
\left.{\delta \mathcal H \over \delta \Phi}\right|_V = \left.{\delta
\mathcal H \over \delta \eta}\right|_V= 0;
\end{align}
for a fixed vortex. The perturbed Hamiltonian in the presence of the
vortex is quadratic in the phase and density variations $\phi, \eta$
and leads to the coupled equations:
\begin{align}\label{qpeoms}
-{\hbar \over m_0} \dot \phi &=
 \left.{\delta^2 \mathcal H \over
\delta \Phi\delta \eta}\right|_V \phi +
\left.{\delta^2 \mathcal H \over \delta \eta^2}\right|_V\eta \\
{\hbar \over m_0} \dot \eta &= \left.{\delta^2 \mathcal H \over
\delta \Phi\delta \eta}\right|_V \eta + \left.{\delta^2 \mathcal H
\over \delta \Phi^2}\right|_V\phi \nonumber
\end{align}
A trivial solution to the perturbed Hamiltonian can be found by
taking the gradient of the original vortex equations (\ref{vortex}):
\begin{align}
\nabla \left.{\delta \mathcal H \over \delta \eta}\right|_V &=
\left.{\delta^2 \mathcal H \over \delta \Phi\delta \eta}\right|_V \nabla \Phi_V +
\left.{\delta^2 \mathcal H \over \delta \eta^2}\right|_V\nabla \rho_V =0\\
\nabla \left.{\delta \mathcal H \over \delta \Phi}\right|_V &=
\left.{\delta^2 \mathcal H \over \delta \Phi\delta \eta}\right|_V
\nabla \rho_V + \left.{\delta^2 \mathcal H \over \delta
\Phi^2}\right|_V\nabla \Phi_V =0
\end{align}
where $\Phi_V$ and $\rho_V$ are the 'texture' solutions in the
presence of the vortex. Comparing with the equations of motion
resulting from the perturbed Hamiltonian, we see that the derivative
of the original vortex profile satisfies them at zero frequency. The
zero modes are then $\phi_0=\nabla \Phi_V\cdot \hat n, \eta_0 =
\nabla \rho_V \cdot \hat n$ where the derivatives have been
projected onto an arbitrary direction $\hat n$.

The zero mode can also  be found by considering a small translation
of the vortex.  Expanding the vortex solution about a shifted
position ${\bf r} +\delta {\bf r}$, we have:
\begin{align}
\Phi_V({\bf r}+ \delta{\bf r})& \approx \Phi_V({\bf r}) +
\delta {\bf r} \cdot \nabla \Phi_V({\bf r}) \\
&= \Phi_V({\bf r}) - \delta {r} \sin(\theta-\theta_d)
{1\over r} \partial_\theta\Phi_V({\bf r}) \nonumber \\
\rho_V({\bf r}+ \delta{\bf r})& \approx \rho_V({\bf r}) +
\delta {\bf r} \cdot \nabla \rho_V({\bf r})\\
&=\rho_V({\bf r}) + \delta {r} \cos(\theta-\theta_d) \partial_r
\rho_V({\bf r}) \nonumber
\end{align}
The zero mode corresponds to the prefactor of the small translation
$\delta r$, ie., the zero mode generates translations of the vortex,
and indeed, the zero mode degrees of freedom are equivalent to the
vortex degrees of freedom.

There are two key observations that follow from this discussion.
First, and rather obviously, we must now exclude the zero modes when
defining the quasiparticle excitations. Thus, the quasiparticle
wave-functions must now be redefined so as to be orthogonal at all
times to the zero modes, so that as the vortex moves, the
quasiparticle wave-functions must continuously adapt to the changing
position of the vortex (indeed, they are excluded from the vortex
core, and phase-shifted far from the vortex). Second, although there
will be an interaction between the vortex and the new perturbed
quasiparticles, this interaction cannot have any term linear in the
quasiparticle variables. This is because the vortex soliton is
itself a minimum action solution to the equations of motion, and so
any fluctuations about this solution (corresponding to the perturbed
quasiparticles) are at lowest order quadratic in the fluctuation
variables.

These points are familiar in the discussion of quantum solitons in
$1+1$-d field theories\cite{Raja}; a well-known example is the
quantum Sine-Gordon model. However $1+1$-dimensional field theories
are in many ways rather unique, and the standard belief for a long
time has been that quantum soliton problems in higher dimensions
were intractable. In fact this is not the case\cite{TS12}; however,
the vortex problem does bring in some interesting new features,
notably:

(i) unlike most of the interesting $1+1$-dimensional models, the
quasiparticle spectrum is gapless here. This, along with the
long-range nature of the interaction between vortices and
quasiparticles, emphasizes the infra-red part of their coupling -
indeed we expect to find divergences in the coupling to the
unperturbed quasiparticles (which would mean that any perturbative
or diagrammatic expansion in powers of this coupling would be at
best unreliable, at worst meaningless). However, as we shall see,
the coupling to the perturbed quasiparticles is not IR divergent.

(ii) the perturbed quasiparticles differ from the unperturbed plane
wave quasiparticles not only in the spatial form of their
wave-function - they are also now chiral excitations, with angular
momentum defined relative to the vortex position.

The upshot of all of this is that we must now distinguish between
the original quasiparticles, described by the field variables
$\phi({\bf r},t)$ and $\eta({\bf r},t)$, and the new perturbed
quasiparticles. We can describe the low-energy dynamics of the
system by defining these new variables as excitations about the
vortex texture, ie., we write:
\begin{align}
\Phi({\bf r},t) &= \Phi_V({\bf r} - {\bf R}(t)) + \tilde{\phi}({\bf r},t|{\bf R}(t))\\
\rho({\bf r},t) &= \rho_V({\bf r} - {\bf R}(t)) + \tilde{\eta}({\bf
r},t|{\bf R}(t))
\end{align}
where the notation makes clear that the quasiparticles are tied to
the vortex position. We can then write $\tilde\phi$ and $\tilde\eta$
in cylindrical components centered at the instantaneous position
${\bf R}(t)$ of the vortex, which we write as
\begin{align}
\tilde{\phi}(r,\theta,t) &= \tilde{\phi}_{l{ k}}(r) \sin(\omega_{ k} t + l\theta) \\
\tilde{\eta}(r,\theta,t) &= -\tilde{\eta}_{l{ k}}(r) \cos(\omega_{
k} t + l\theta)\nonumber
\end{align}
We see that the time and angular dependence cannot be separated in
the perturbed quasiparticles; they are now chiral modes tied to the
background vortex.

From this discussion one might imagine that we can now completely
forget about the original plane wave quasiparticles. In an isolated
system this would indeed be the case. However in experiments one can
do something rather interesting, which is to inject 'external' plane
wave quasiparticles - in effect, one can irradiate the vortex with
an external quasiparticle wind. These quasiparticles are {\it not}
orthogonal to the vortex 'zero mode' wave-function, and they will
interact linearly with it. Below, we discuss the experimental
implications of this point.


\section{Vortex-Quasiparticle interaction}

Formally, we may now write the expansion of the superfluid action in
terms of the perturbed (tilded) quasiparticles in the form:
\begin{align}
S = \tilde S_v^0[{\bf R}(t)]+ \tilde S_{qp}[\{\tilde \phi,\tilde
\eta\}] + \Delta S_{int}^{(2)}[\{\tilde \phi,\tilde \eta\}]
\end{align}
where the zero mode is accounted for in ${\bf R}(t)$; here $\tilde
S_v^0[{\bf R}(t)]$ is the vortex action and $\tilde S_{qp}[\{\tilde
\phi,\tilde \eta\}]$ is the quasiparticle action (both written in
terms of perturbed quasiparticles), and $\Delta
S_{int}^{(2)}[\{\tilde \phi,\tilde \eta\}]$ is the interaction term
(where the superscript indicates that it is quadratic in the
perturbed quasiparticle variables). Rather than give a lengthy
discussion of how $\Delta S_{int}^{(2)}[\{\tilde \phi,\tilde
\eta\}]$ is calculated, let us instead discuss the result, which can
be portrayed in terms of the Feynman diagrams for the final form of
the vortex-quasiparticle interaction. One may give these results
either in terms of unperturbed 'external' quasiparticles, or in
terms of the perturbed (tilded) quasiparticles - here we focus on
the perturbed quasiparticles.

The key question is of course to understand the form of the
interaction between the vortex and the perturbed quasiparticles,
which is incorporated in $\Delta S_{int}^{(2)}[\{\tilde \phi,\tilde
\eta\}$. To give an intuitive feel for this interaction, we sketch
here the form of the effective field theory which describes it, in
diagrammatic terms.

In a large system, the quasiparticle propagators are the same for
perturbed or unperturbed quasiparticles. We define the quasiparticle
matrix propagator $G_{km}^o(\omega)$ by
\begin{align}
G_{km}^o(\omega)\left( \begin{array}{cc}
{\hbar \rho_s}k^2& m_0 \omega \\
m_0 \omega&  {m_0^2\over \hbar \rho_s^2 \chi} \end{array} \right) =
1
\end{align}

In the same way one may define a propagator for the vortex itself,
starting from $\tilde S_v^0[{\bf R}(t)]$. Consider now the diagrams
for the interaction between the quantum zero mode and the perturbed
quasiparticles. One of the vertices involved is shown in Fig. 1.

\begin{figure}
\vspace{0.4cm}
\includegraphics[width=0.5\textwidth]
{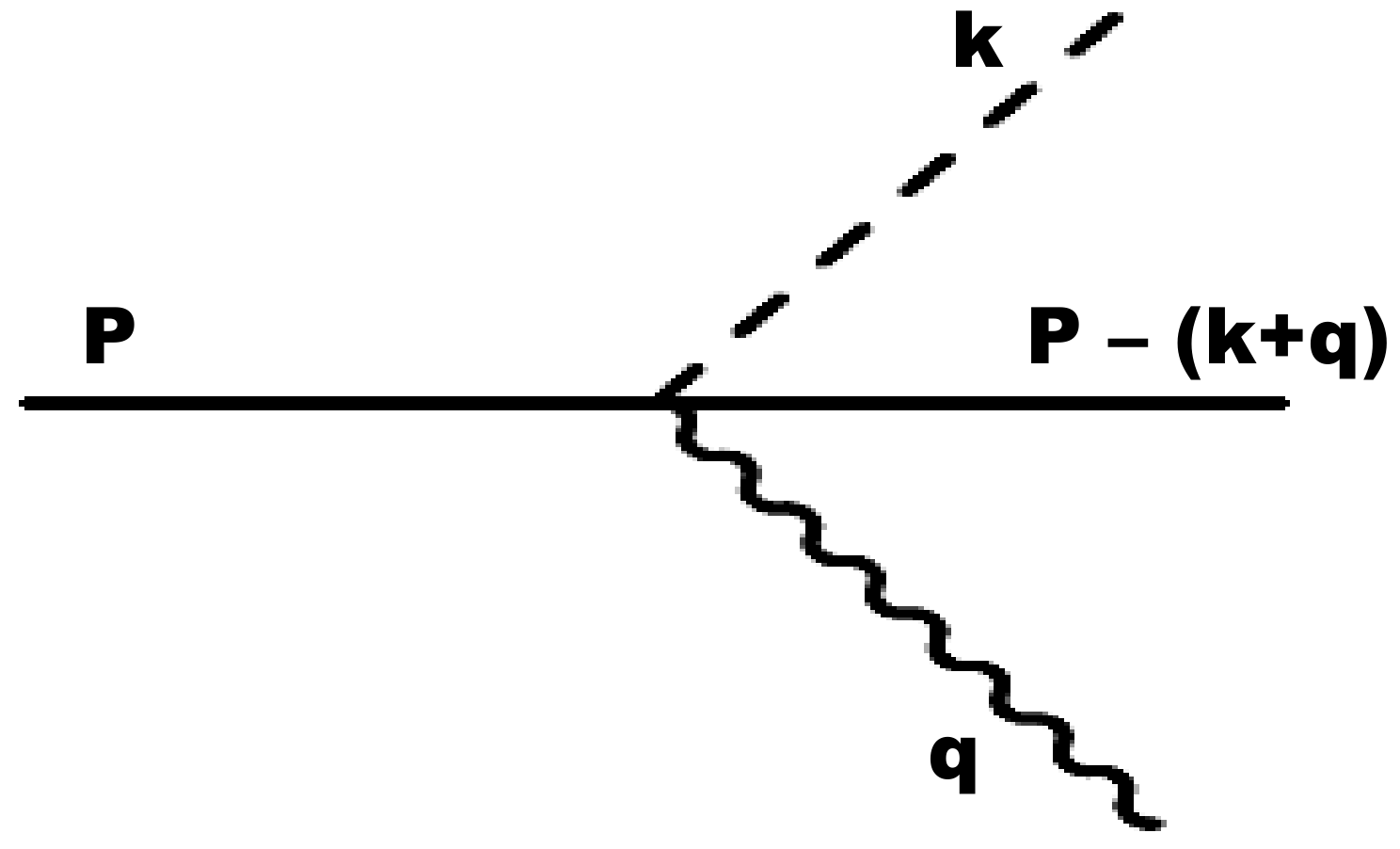} \caption{Interaction between a vortex
zero mode and a pair of perturbed (tilded) quasiparticles in a Bose
superfluid. The dashed and wavy lines indicate quasiparticles of
opposite chirality, and the momenta of each excitation are also
shown. }
 \label{fig:VortInt}
\end{figure}

The expression for the total coupling $\Lambda_{kq}^{\sigma l}$,
between the vortex and a pair of quasiparticles having momenta $k$
and $q$ respectively, takes the following form:
\begin{align}
 \label{intmatrix}
\Lambda_{kq}^{\sigma l} = \int {dr\over 2m_o} \; \Big[ r&(\tilde
\phi_{lk}\partial_r \tilde\eta_{l+\sigma,
q}+\tilde\eta_{lk}\partial_r \tilde\phi_{l+\sigma, q}) \\
&+ \sigma (l+
\sigma) (\tilde\phi_{lk} \tilde\eta_{l+\sigma, q}+\tilde\eta_{lk}
\tilde\phi_{l+\sigma, q}) \Big]\nonumber
\end{align}
This interaction is zero unless $\sigma = \pm 1$: the moving vortex
thus only couples the renormalized modes $\tilde{\phi}_{l{\bf k}}$
and $\tilde{\eta}_{l'{\bf q}}$ to each other if $|l-l'|=1$,
transferring angular momentum $\hbar \sigma = \hbar (l'- l)$ between
them. Because of the long-range vortex field, we focus on the
'long-wavelength' regime where $k a_o \ll 1$. In this regime, only
the transitions between $l=0$ and $l=\pm1$ contribute to linear
order in $a_ok$. Using the anti-symmetry of $\Lambda$ under exchange
of the initial state $(k,l)$ and the final state $(q,l+\sigma)$,
viz., $\Lambda_{kq}^{\sigma l} = -\Lambda_{qk}^{-\sigma, l+\sigma}$,
we can fully express the total coupling by the $l=0$ term, to get:
\begin{align}
 \label{Lambda0}
\Lambda_{kq}^{\sigma 0} &={k+q \over 4\sqrt{kq}}\delta(k-q) \\
& + {a_o
\over 4} \left\{ \begin{array}{ll} {\sigma \over 2}\left({k\over
q}\right)^{3\over 2} +
\sigma\sqrt{kq}{k\over q(k-q)} & \mbox{if $ k<q$} \\
{k+q \over 16\sqrt{kq}}a_oq +{\sigma \over 2} \left({q\over
k}\right)^{1\over 2} + \sigma\sqrt{kq}{q\over k(k-q)} & \mbox{if
$q\leq k$}
\end{array} \right. \nonumber
\end{align}
valid for $q_v=1$.

The form of this interaction is interesting. As noted above, it is
not IR divergent - this is because the perturbed quasiparticle
wave-functions have adjusted to the vortex zero mode wave-function.
Nevertheless it is not analytic about the zero momentum point, and
part of it changes sign with the angular momentum transfer $\sigma$.
Insofar as one believes the long-wavelength description of the Bose
superfluid, the result is exact in the long wavelength limit.

There have of course been many attempts in the past to derive the
form of the interaction between a vortex and the quasiparticles in a
Bose superfluid, and it is useful to compare the result
(\ref{Lambda0}) with some of the forms found in previous work on
this problem. These fall mainly into 2 categories. The first set of
calculations attempts to calculate a scattering amplitude for
quasiparticles interacting with a classical vortex potential. Such
calculations automatically yield a quadratic coupling, ie., a
coupling to quasiparticle pairs. However, the long-range nature of
the vortex field creates infra-red divergences in this scattering
amplitude, which require careful discussion \cite{sonin97,fetter69};
moreover, the potential also carries an effective Aharonov-Bohm flux
\cite{sonin97,stone00}. It actually turns out to be quite difficult
to compare the details of such calculations with the results given
here, mainly because (i) almost all of these calculations deal with
the scattering of plane wave excitations off a static vortex (with
no recoil); and (ii) the vortex itself is not treated
quantum-mechanically.

A second class of calculations employs a Hamiltonian of form
\cite{thouless94}
\begin{align} \label{TAN94}
H = {1 \over 2M_v} & [{\bf P} - q_v {\bf A}({\bf r}]^2 \\
& + \sum_{\bf k}
 \left(c_{\bf k} {\bf q}_{\bf k} \cdot {\bf R} + {1 \over 2 m_{\bf k}} \big[ {\bf p}_{\bf k}^2
 + m_{\bf k} \omega_{\bf k}^2 {\bf q}_{\bf k}^2 \big] \right) \nonumber
\end{align}
where the vector potential ${\bf A}$ yields a `field' $\nabla \times
{\bf A}({\bf r}) = \pi \hbar \rho_s \hat{\bf z}$, which is
responsible for the Magnus force. This Hamiltonian takes the
Feynman-Vernon/Caldeira-Leggett form \cite{feynman63}, with
couplings $c_{\bf k}$ to quasiparticle coordinates $\{ {\bf q}_{\bf
k} \}$ (having conjugate momenta $\{ {\bf p}_{\bf k} \}$),  which
are linear in the $\{ {\bf q}_{\bf k} \}$. As noted above, such a
linear interaction {\it does} exist between the vortex and the {\it
unperturbed} plane-wave quasiparticles. However, as we have already
explained, no linear interaction to correctly orthogonalized
quasiparticles can exist.

Note however that this does not stop us from writing down a
Hamiltonian like (\ref{TAN94}) containing a linear interaction
between the vortex and {\it pairs} of quasiparticles -- such forms
have been employed in other cases involving quasiparticle-soliton
interactions. However the interaction now depends on the momenta of
both quasiparticles, and is usually strongly temperature-dependent;
considerable care is needed to evaluate it. In our view the only
reliable way to carry out such a manoeuvre is to first derive the
interaction between the vortex and the true orthogonalized
quasiparticles, as above, and then from this derive the form of the
interaction to a bath of effective oscillators.

As shown in ref.\cite{TS12}, one can in fact carry though a fully
non-perturbative derivation of the time dynamics of the vortex
system, by dealing directly with the superfluid action, which
incorporates the infra-red convergent interaction (\ref{Lambda0})
between the vortex and the perturbed quasiparticle field. By then
integrating out the quasiparticles one finds an equation of motion
for the vortex reduced density matrix - there is no need to deal
directly with the vortex scattering problem at all.


\section{Equations of Motion}

As we noted at the beginning of this paper, the classical
description of a vortex involves an equation of motion for the
classical coordinate ${\bf R}_v(t)$. As we will discuss below, the
correct classical vortex equation of motion turns out to be
\begin{align}
 \label{eom1}
M_v \ddot {\bf R}_v - {\bm f}_M - {\bm f}_{qp} - {\bf f}_{ac}(t)
\;=\; {\bf f}_{fl}^{(cl)}(t)
\end{align}
where ${\bf f}_{ac}(t)$ is some driving force, $M_v$ is the vortex
mass, ${\bm f}_M = \rho_s {\bm \kappa} \times (\dot {\bf R}_v - {\bm
v}_s)$ is the Magnus force for a vortex with circulation ${\bm
\kappa} =\hat {\bf z}h/m$, and the quasiparticle force ${\bm
f}_{qp}$ is
\begin{equation}
{\bm f}_{qp} \;=\; D_o ({\bm v}_n - \dot {\bf R}_v) + D_o^{\prime}
\hat {\bf z} \times ({\bm v}_n - \dot {\bf R}_v)
 \label{f-QP}
\end{equation}
in which the longitudinal drag $D_o(T)$ and the transverse term
$D_o^{\prime}(T)$ depend strongly on the temperature $T$. The
classic discussion of Iordanskii  yields
\begin{equation}
D_o^{\prime}(T) = - \kappa \rho_n(T)
 \label{iordF}
\end{equation}

Finally, ${\bf f}_{fl}^{(cl)}(t)$ is a the classical limit of a
'fluctuational noise' term ${\bf F}_{fl}(t)$, whose behaviour is
defined by its correlator $\chi_{ij}(t-t',T) = \langle F^i_{fl}
(t,T) F^j_{fl}(t',T)\rangle$. In the classical regime this
correlator takes the form
\begin{equation}
\chi_{ij}(t-t',T) \longrightarrow \chi_{ij}^{(cl)}(t-t', T)
\sim \chi_o^{\parallel} (T) \delta_{ij} \delta(t-t')
 \label{chi-cl}
\end{equation}
ie., it is entirely longitudinal, and displays Markovian white noise
- we discuss the temperature dependence below.

Now eqtn. (\ref{eom1}) is in fact the original HVI
equation\cite{HallV,iord}, but with an added inertial mass term and
a longitudinal noise term. It is actually a local (in spacetime)
equation, ie., it can be written in the form $\hat{\cal L}(t) {\bf
R}_v(t) = f(t)$, where $\hat{\cal L}(t)$ is a local differential
operator acting at time $t$, involving forces and an inertial term
which act on ${\bf R}_v(t)$ at time $t$ only. There is actually no
reason why the classical dynamics need to be local - one could
easily have, eg., 'memory' terms in the dissipation of form $\int
dt' \Gamma_{ij}(t-t') \dot{R}_v^j(t')$, and quite generally one
could have an equation of form $\hat{\cal L}(t,t'){\bf R}_v(t')$,
where $\hat{\cal L}$ is now some integrodifferential operator
function of $t$ and $t'$. However, we will see that the correlation
times in the classical regime are very short, and a local equation
is a very accurate approximation to the truth.

Consider now the quantum dynamics. This has to be written in terms
of an equation of motion for the reduced density matrix $\bar{\bm
\rho}_v({\bf R}, {\bf R'}, t)$. Now the range of possible different
forms for an equation of motion for $\bar{\bm \rho}_v({\bf R}, {\bf
R'}, t)$ is very large. Quite generally we might expect the equation
to be non-local in the variables ${\bf R}, {\bf R'}$, and $t$.
Moreover, there is no reason to assume that we will be able to write
the equation of motion in terms of simple forces, which are a
classical notion of limited applicability in quantum mechanics.

It is then refreshing to find that the actual time dynamics of
$\bar{\bm \rho}_v({\bf R}, {\bf R'}, t)$ do assume a fairly simple
form, even in the quantum regime. For a discussion of experiments it
is convenient to transform the equations of motion to the frequency
domain, and we discuss this in the next section. But it is also of
interest to look at them in the real time domain, which we do here.
The derivation of these results is described in refs.
\cite{TS12,TSlong}. The key assumption is that we can make a
Born-Oppenheimer expansion, assuming the vortex velocity is small
compared to the sound velocity in the superfluid. For the derivation
of the forces acting on the vortex this expansion is perfectly well
behaved, and we can thus have confidence in the results. However the
derivation of the effective mass is more subtle and the
Born-Oppenheimer expansion misses the 'radiation reaction' terms
(which also exist classically). Consequently the Born-Oppenheimer
expansion yields a hydrodynamic mass $M_v^o$ for the vortex, without
frequency-dependent corrections or higher time derivatives (eg.,
terms proportional to $\dddot{\bf R}_v$).

The results for the vortex dynamics can be written in terms of an
equation of motion for $\bar{\bm \rho}_v({\bf R}, {\bf R'}, t)$, but
it is more illuminating here to give them in terms of the dynamics
of the arguments of $({\bf R}, {\bf R'}$. We define the sum and
difference variables ${\bf R}_v \;= \;{1\over 2}({\bf R}+{\bf R}')$
and ${\bm \xi} \;= \;{\bf R}-{\bf R}'$, so that ${\bf R}_v(t)$
denotes a 'centre of mass' coordinate for the vortex, and ${\bm
\xi}(t)$ a 'quantum fluctuation' coordinate about the centre of mass
coordinate.

One can then write the results in terms of equations of motion for
these 2 variables. For the centre of mass coordinate one gets an
equation which can be written as
\begin{align}
 \label{Rveom}
M_v^o \ddot{\bf {R}}_v(t)& - {\bm f}_M(\dot{\bf {R}}_v) - {\bf
F}_{QP}^{\bf R}[\dot{\bf {R}}_v - {\bf v}_n] = {\bf F}_{fluc}(t)
\end{align}
where ${\bm f}_M(\dot{\bf {R}}_v)$ is again the Magnus force, and
where the new quasiparticle force ${\bf F}_{QP}^{\bf R}$ is now a
non-local {\it functional} of the velocity of the vortex, relative
to the normal velocity. The equation for ${\bm \xi}(t)$ takes a
somewhat similar form:
\begin{align}
 \label{xieom}
M_v^0 \ddot {\bm \xi}(t)& - {\bm f}_M(\dot {\bm \xi} )  - {\bf
F}_{QP}^{\bm \xi}[\dot{\bm \xi}(t)] = 0
\end{align}
where however now ${\bm f}_M(\dot {\bm \xi} )= \rho_s q_v{\bm
\kappa} \times \dot {\bm \xi}(t)$ (ie., this force is like the
Magnus force acting on the relative velocity $( \dot{\bf {R}}_v(t) -
{\bm v}_s)$, except it now acts simply on the `fluctuation velocity'
$\dot {\bm \xi}$); where the 'quasiparticle' term does not depend on
${\bf v}_n$; and where there is no noise fluctuation term.

We do not have space here to look in detail at the equation of
motion for ${\bf \xi}(t)$, but it is physically illuminating to look
at the new quasiparticle force term ${\bf F}_{QP}^{\bf R}[\dot{\bf
{R}}_v - {\bf v}_n]$. To make the connection with the HVI classical
force ${\bm f}_{qp}$ appearing in (\ref{f-QP}), let us write its
quantum generalization as
\begin{equation}
{\bf F}_{QP}^{\bf R}[\dot{\bf {R}}_v - {\bf v}_n] \;=\; {\bf
F}_\|^R[\dot{\bf {R}}_v - {\bf v}_n] \;+\; {\bf F}_\perp^R[\dot{\bf
{R}}_v - {\bf v}_n]
\end{equation}
where the idea is to separate out the two terms that in the
classical limit lead to the drag force and the Iordanski force.

Let us consider in detail the 'parallel' term ${\bf F}_\|^R[\dot{\bf
{R}}_v = {\bf v}_n]$. It is a functional of the prior vortex
velocity parallel to the normal fluid; in fact it takes the
microsopic form:
\begin{align}
{\bf F}_\|^R[\dot{\bf {R}}_v - {\bf v}_n] =  {\hbar\over L_z}
&\sum_{m\sigma kq}
(\Lambda_{kq}^{\sigma m})^2 \Omega_{kq}(n_k - n_q) \\
&\times \int_{t_1}^t ds \; (\dot{\bf {R}}_v(s)- {\bm v}_n) \cos
[\Omega_{kq}(t-s)]\nonumber
\end{align}
where we have assumed a quasi-2d film of thickness $L_z$, and the
frequency $\Omega_{kq}$ is just the difference in energies between
the two quasiparticles that interact with the vortex, ie.,
$\Omega_{kq} = \omega_k - \omega_q$, where we expect that in the
long wavelength regime, $\hbar \omega_k = c_s |{\bf k}|$, where
$c_s$ is the sound velocity. We see that this force has just assumed
a simple 'memory' form, and that its instantaneous value and
direction depend on the previous path traced out by the vortex (more
precisely, the component of that the vortex velocity along that path
that was parallel to ${\bf v}_n$). However there is no requirement
for ${\bf F}_\|^R[\dot{\bf {R}}_v - {\bf v}_n]$ at a given time $t$
to be parallel to ${\bf v}_n$ at the position ${\bf R}_v$ of the
vortex - if ${\bf v}_n$ varies with time or with position, then this
will not in general be the case.

Consider now the behaviour in time of this 'memory term'. To do
this, let us imagine a vortex following a straight line trajectory -
we can then write the parallel force as
\begin{equation}
{\bf F}_\|^R[\dot{\bf {R}}_v - {\bf v}_n] \rightarrow \int^t ds
D_{\parallel}(t-s;T)({\bf v}_n - \dot{\bf {R}}_v(s))
\end{equation}
where the function $D_{\parallel}(t-s;T)$ is a non-local (in time)
generalization of the HVI coefficient $D_o(T)$. Now this function
turns out to depend only on the product $kT(t-s)/\hbar$ (so that its
Fourier transform, as advertised, depends only on the ratio
$\tilde{\Omega} = \hbar \Omega/kT$). In Fig. 2 we show its behaviour
as a function of renormalized time.

\begin{figure}
\includegraphics[width=0.5\textwidth]{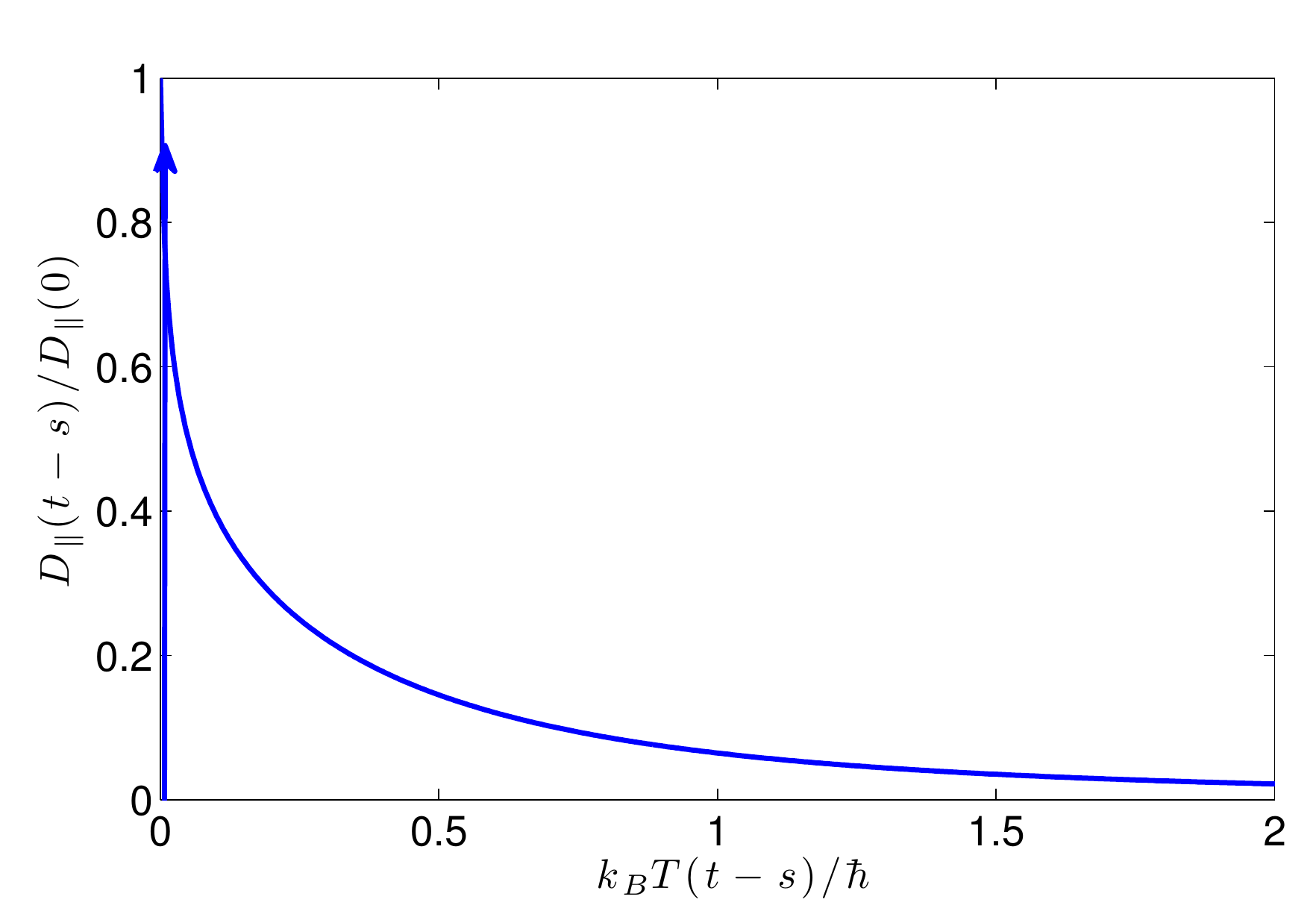}
\caption{The longitudinal damping time integral kernel $D_\|(t-s)$,
shown as a function of the dimensionless variable $k_BT(t-s)/\hbar$:
the coefficient decays roughly as $(t-s)^{-2}$ at  long times, a
relatively slow decay coming from the linear low-frequency behaviour
of $D_\|(\Omega)$. The arrow at the origin denotes a local in time
($\delta$-function) damping contribution.  }
 \label{fig:Dlong-t}
\end{figure}

Suppose we write the longitudinal damping function
$D_{\parallel}(t-s;T)$ as a sum of a local $\delta$-function
contribution and a retarded term. Now at first glance, one's
intuitive expectation is that the $\delta$-function term will simply
be equal to the classical $D_o(T)$ contribution. But this is not
correct. Actually one finds that
\begin{equation}
D_\| (t-s, T) =   {1 \over 16} D_o (T)\; \delta (t-s) \;+\; \delta
D_\| (t-s; T)
 \label{D-t-s}
\end{equation}
where the long-time tail term $\delta D_\| (t-s; T)$ behaves roughly
as $\sim (t-s)^{-2}$. Thus as one tends to low temperatures or short
times, such that $k_BT(t-s)/\hbar \ll 1$, one is left with the
'quantum contribution' to the longitudinal damping - but this is 16
times smaller than $D_o(T)$, and moreover, it behaves like a local
term! It is only at high temperatures or long times, when one
integrates over all of $D_\| (t-s, T)$, that the full contribution
to the classical coefficient $D_o(T)$ is recovered.

From these remarks one sees that the crossover between the classical
HVI equations and the fully quantum regime is going to be an
interesting one. We have no space here to outline the behaviour of
the other terms in the equation of motion - suffice it to say that
both the transverse quasiparticle force and the fluctuation force
have non-local memory terms (although the correction to the
Iordanskii force in the transverse term turns out to be very small).
Remarkably, once we have made the full crossover to the quantum
regime (ie., where $kT(t-s)/\hbar \ll 1$, or where $\hbar \Omega \ll
kT$, one actually ends up again with a local equation of motion for
the vortex, this time coming only from the $\delta$-function terms
in the various memory kernels. This equation is
\begin{align}
 \label{eom1b-Q}
M_v \ddot {\bf R}_v - {\bf f}_M - {\bm F}_{qp}^{(Q)} - {\bf
F}_{ac}(t) \;=\; {\bf F}_{fluc}^{(Q)}(t, T)
\end{align}
where the quasiparticle force ${\bf F}_{QP}^{(Q)}$ is given by
\begin{equation}
{\bm F}_{QP}^{(Q)} \;=\; {1 \over 16} D_o ({\bm v}_n - \dot {\bf
R}_v) + D_o^{\prime} \hat {\bf z} \times ({\bm v}_n - \dot {\bf
R}_v)
 \label{f-QP-Q}
\end{equation}
and where the fluctuation correlator is again Markovian and entirely
longitudinal:
\begin{equation}
\chi_{ij}^{(Q)}(t-s, T) \;=\; {\zeta(5)\over 4\zeta(4)}
\chi_o^{\parallel}(T) \delta_{ij} \delta(t-s)
 \label{chi-ij-Q}
\end{equation}
Thus the equation of motion in the extreme quantum regime has
exactly the same form as the classical HVI equation, but with quite
different coefficients (except for the Magnus and Iordanskii terms,
which have exactly the same coefficients).

To summarize - one finds that the actual equations of motion for a
quantum vortex are rather complicated except in the extreme
classical regime, where they reduce to the standard HVI equation,
and in the extreme quantum regime, where they reduce to equation
(\ref{eom1b-Q}). The intervening crossover regime is expected to
show quite different behaviour from either of the two limiting
cases.


\section{Conclusions, and Remarks on Experiments}

Condensed matter systems are populated by 3 different kinds of
quantum excitation - extended quasiparticle modes, localized modes
such as spins, or defects, and quantum solitons. It is obviously of
great importance to understand how these different excitations
interact, and the debate over the nature of vortex-quasiparticle
interactions has assumed a central importance in the theory of
superfluids over the years. In this work and in
refs.\cite{TS12,TSlong} we present what we think is a solution to
this problem, obtained by extending the theory beyond the purely
classical regime. We should however note some of the limitations of
this work. First, it is only valid in the long wavelength limit - we
ignore higher-order interquasiparticle interactions, and excitations
like rotons in superfluid $4$He (in $^4$He this confines us to $T <
0.6-0.7~K$). It also means that we cannot deal with the crossover in
the vortex flow field between the 'near' regions where normal fluid
viscosity can be neglected, and the far region where the viscosity
controls the flow. Thouless et al.\cite{ThV01}, have shown there are
subtle problems involved in this crossover, and in fact we believe
that the question of the total circulating normal fluid to be found
around a vortex still needs to be settled. A second limitation is
that the assumption of a slow vortex and a Born-Oppenheimer
expansion, so that we cannot capture all terms contributing to the
inertial forces on a vortex. Finally, the work here describes
vortices in a Bose superfluid - vortices in Fermi superfluids have
to be dealt with separately.

It is nevertheless interesting to speculate on how one might address
the quantum regime in experiments. One needs high frequency vortex
motion. The obvious candidates for experiments then include (i)
vortex tunneling experiments (where the bounce frequency is very
high) (ii) experiments on the dynamics of single vortices in
2-dimensional 'pancake' cold BEC gases - this problem is discussed
by Cox and Stamp\cite{cox12}; and (iii) very low-$T$ experiments on
turbulence (where vortex motions can be extremely rapid). The
detailed theory of such experiments remains an interesting
challenge.


\begin{acknowledgements}

This work was supported by funding from NSERC, from PITP, and from
CIFAR. The work benefited greatly from discussions at various times
with (and encouragement from) David Thouless; we also thank Bill
Unruh and Gordon Semenoff for useful comments.

\end{acknowledgements}


\begin{thebibliography}{99}

\bibitem{vinen56}    W.F. Vinen, Nature {\bf 181}, 1524 (1958).

\bibitem{HallV}      H. E. Hall and W. F. Vinen, {\it Proc. R. Soc. A} \textbf{238}, 204, 215 (1956).

\bibitem{iord}       S.V. Iordanskii, {\it Ann. Phys. (N.Y.)} \textbf{29}, 335 (1964).

\bibitem{anglin07}   D. J. Thouless and J. R. Anglin, {\it Phys. Rev. Lett.} \textbf{99}, 105301 (2007).

\bibitem{TAN96}      D. J. Thouless, P. Ao, and Q. Niu, {\it Phys. Rev. Lett.} \textbf{76}, 3758 (1996).

\bibitem{sonin97}    E. B. Sonin, {\it Phys. Rev. B} \textbf{55}, 485 (1997).

\bibitem{fetter64}   A.~L. Fetter, Phys. Rev. {\bf 136}, A1488 (1964)

\bibitem{wexthou98}  C.~Wexler and D.~J. Thouless, Phys. Rev. {\bf B58}, R8897 (1998)

\bibitem{stone00}    M. Stone, Phys. Rev. {\bf B61}, 11780 (2000)

\bibitem{TS12}       L. Thompson and P. C. E. Stamp, {\it Phys. Rev. Lett.} \textbf{108}, 184501, (2012).

\bibitem{ThV01}      D.J. Thouless et al., Phys. Rev. {\bf B63}, 224504 (2001)

\bibitem{TSlong}     See L. Thompson, PCE Stamp, in preparation; and L. Thompson, PhD thesis, University of
                     British Columbia (2011)

\bibitem{Raja}       R. Rajaraman, ``{\it Solitons and Instantons}'' (Elsevier,
                     1987); T.D. Lee, ``{\it Particle Physics and Introduction to Field
                     Theory}'', Ch. 7 (Harwood, 1981)

\bibitem{ons56}      O. Penrose, L Onsager, Phys. Rev. {\bf 104}, 576 (1956); CN Yang, Rev. Mod. Phys.
                     {\bf 34}, 694 (1962); JS Langer, Phys. Rev. {\bf 167}, 183 (1968).


\bibitem{fetter69}   A.L. Fetter, Phys. Rev. {\bf 186}, 128 (1969)


\bibitem{thouless94} P. Ao, D.J. Thouless, Phys. Rev. Lett. {\bf 72}, 132 (1994);
                     Q. Niu, P. Ao, D.J. Thouless, Phys. Rev. Lett. {\bf 72}, 1706 (1994)

\bibitem{feynman63}  R.~P. Feynman and F.~L. Vernon, Ann. Physics, {\bf 24}, 118 (1963);
                     A. O. Caldeira and A. J. Leggett, Ann. Phys. {\bf 149}, 374 (1983).


\bibitem{cox12}      T Cox, PCE Stamp, J. Low Temp. Phys., this issue (2012)







\end{thebibliography}
\end{document}